\documentclass[journal=nalefd,manuscript=article]{achemso}

\usepackage[version=3]{mhchem} 
\usepackage{color}
\usepackage{ulem}

\newcommand{\etal}{{\em et al}}
\newcommand{\lsmo}{{L\lowercase{a}$_{0.67}$S\lowercase{r}$_{0.33}$M\lowercase{n}O$_3$}}

\newcommand{\pto}{{P\lowercase{b}T\lowercase{i}O$_3$}}
\newcommand{\pzt}{{P\lowercase{b}Z\lowercase{r}$_{0.2}$T\lowercase{i}$_{0.8}$O$_3$}}
\newcommand{\sro}{{S\lowercase{r}R\lowercase{u}O$_3$}}
\newcommand{\sto}{{S\lowercase{r}T\lowercase{i}O$_3$}}

\author{C\'eline Lichtensteiger}
\email{celine.lichtensteiger@unige.ch}
\affiliation{DPMC - University of Geneva, 24 Quai Ernest Ansermet, CH - 1211 Geneva 4, Switzerland}
\author{St\'ephanie Fernandez-Pena}
\affiliation{DPMC - University of Geneva, 24 Quai Ernest Ansermet, CH - 1211 Geneva 4, Switzerland}
\author{Christian Weymann}
\affiliation{DPMC - University of Geneva, 24 Quai Ernest Ansermet, CH - 1211 Geneva 4, Switzerland}
\author{Pavlo Zubko}
\email{p.zubko@ucl.ac.uk}
\affiliation{London Centre for Nanotechnology and Department of Physics and Astronomy, University College London, 17-19 Gordon Street, London WC1H 0HA, UK}
\altaffiliation{Previous address: DPMC - University of Geneva, 24 Quai Ernest Ansermet, CH - 1211 Geneva 4, Switzerland}
\author{Jean-Marc Triscone}
\affiliation{DPMC - University of Geneva, 24 Quai Ernest Ansermet, CH - 1211 Geneva 4, Switzerland}
\phone{+41 (0) 22 379 32 45}
\fax{+41 (0) 22 379 68 69}

\title{Tuning of the depolarization field and nanodomain structure in ferroelectric thin films}


\begin{document}

\begin{tocentry}
\begin{center}
\includegraphics{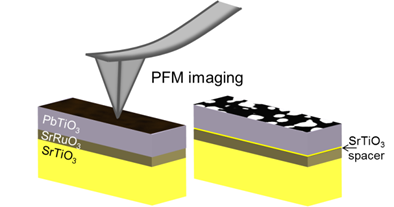}
\end{center}
\end{tocentry}

\begin{abstract}

The screening efficiency of a metal-ferroelectric interface plays a critical role in determining the polarization stability and hence the functional properties of ferroelectric thin films. Imperfect screening leads to strong depolarization fields that reduce the spontaneous polarization or drive the formation of ferroelectric domains. We demonstrate that by modifying the screening at the metal-ferroelectric interface through insertion of ultrathin dielectric spacers, the strength of the depolarization field can be tuned and thus used to control the formation of nanoscale domains. Using piezoresponse force microscopy, we follow the evolution of the domain configurations as well as polarization stability as a function of depolarization field strength.
\end{abstract}

The behavior of ferroelectric thin films is to a large degree determined by the screening of their spontaneous polarization. In ferroelectric capacitors, this screening is provided by the free charges from the metallic electrodes. However, even for ideal, structurally perfect metal-ferroelectric interfaces, this screening is not complete as the screening charges will spread over a small but finite distance, giving rise to a non-zero effective screening length $\lambda_{\rm eff}$ and a corresponding depolarization field inside the ferroelectric~\cite{Mehta-JAP-1973, Junquera-NAT-2003, Dawber-JPCM-2003,Lichtensteiger-Wiley-2011}. This depolarization field increases as the inverse of the thickness of the ferroelectric film and has a detrimental effect on the homogeneous polarization state often favored in applications, leading to either a reduction of the spontaneous polarization or formation of ferroelectric domains \cite{Lichtensteiger-PRL-2005,Nagarajan-JAP-2006, Lichtensteiger-APL-2007}. \pzt\ capacitors with SrRuO$_3$ electrodes, for example, undergo a crossover from a state of uniform polarization to a polydomain state as their thickness is reduced below approximately 150 \AA ~\cite{Nagarajan-JAP-2006}. A similar crossover was observed in \pto\ films with \lsmo\ bottom electrodes\cite{Lichtensteiger-APL-2007}. In both cases, the origin of the polydomain state was attributed to the depolarization field arising from incomplete screening of the spontaneous polarization in the monodomain state. Note that throughout this article, the depolarization field we refer to is the field that would be generated by a uniform polarization in a capacitor with a finite screening length and that serves as the driving force for the domain formation; it is not the stray field associated with the actual polydomain ground state.

The effective screening length has been calculated theoretically for a number of metal-ferro\-electric interfaces \cite{Stengel-NatMat-2009, Junquera-NAT-2003, Gerra-PRL-2006}, and yet, 
despite being frequently invoked to explain various observations\cite{Dawber-JPCM-2003,Kim-PRL-2005,Nagarajan-JAP-2006}, there have been few attempts to systematically tune it or quantify it experimentally \cite{Lichtensteiger-PRL-2005}. The effective screening length is an intrinsic property of a specific metal-ferroelectric interface and can be reduced by a suitable choice of the metallic electrode~\cite{Stengel-NatMat-2009, Saad-JPCM-2004, Chang-AdvMat-2009}. One can also change the screening efficiency of the electrode in a controlled way by inserting dielectric layers between the  electrodes and the ferroelectric to modify the depolarization field inside the ferroelectric layer. Such interface modification has previously been explored in the context of increasing the energy storage capacity of ferroelectric capacitors~\cite{McMillen-APL-2012} and modifying the built-in dipole at metal-ferroelectric interfaces \cite{Lu-AdvMater-2012}. Here, we show how the insertion of a dielectric \sto\ spacer can be used to increase the depolarization field and hence induce ferroelectric nano\-domains in thin \pto\ films. Using piezoresponse force microscopy (PFM), we study the evolution of the domain configuration with \pto\ and \sto\ layer thicknesses and show that the stability of written domains depends on the thickness of the dielectric \sto\ spacers.

The oxide heterostructures used for this study were grown epitaxially on TiO$_2$-terminated (001)-oriented  \sto\ substrates using off-axis radio-frequency magnetron sputtering and consisted of a 22~nm-thick SrRuO$_3$ bottom electrode, followed by a thin \sto\ spacer, a 10--50~nm-thick \pto\ film and a thin \sto\ capping layer. The thicknesses of the top and bottom \sto\ layers were independently varied between 0 (no spacer) and 10 unit cells (uc) i.e. 0 and 4 nm respectively. The deposition conditions are given in Reference [~\bibnote{ 
SrRuO$_3$ was deposited at 640$^\circ$C in 100 mTorr of O$_2$/Ar mixture of ratio 3:60 using a power of 80 W. The \pto\ and \sto\ layers were deposited at 530$^\circ$C in 180 mTorr of 29:20 O$_2$/Ar mixture using a power of 60 W. A Pb$_{1.1}$TiO$_3$ target with a 10\% excess of Pb was used to compensate for the Pb volatility, while stoichiometric targets were used for \sro\ and \sto .}].

Using PFM~\cite{Hidaka-APL-1996,Alexe-Book-2004,Gruverman-JVacSciTechnolB-1996}, we investigated the intrinsic polarization configuration of these hete\-ro\-struc\-tures at room temperature, focusing on the effect of the \pto\ and \sto\ thickness and the change in the degree of screening. 
Amplitude and phase signals were recorded using an Asylum Research Cypher atomic force microscope (AFM) operating in dual resonance tracking (DART) mode~\cite{Gannepalli-Nanotec-2011}. 
The very small lattice mismatch between the \sto\ ($a=3.905$\AA) and the $a$=$b$=3.9045\AA\ axes of tetragonal \pto ~\cite{Landolt-Bornstein-2000} allowed all the films to be epitaxially strained to the substrate forcing the tetragonal $c$-axis to point out-of-plane. 
The polarization can therefore take either of two orientations: {\it up} (pointing away from the substrate) or {\it down} (pointing towards the substrate), leading to 180$^\circ$ domains when regions with these two different orientations develop in a film.

\begin{figure}[!htb]
\includegraphics[width=\columnwidth]{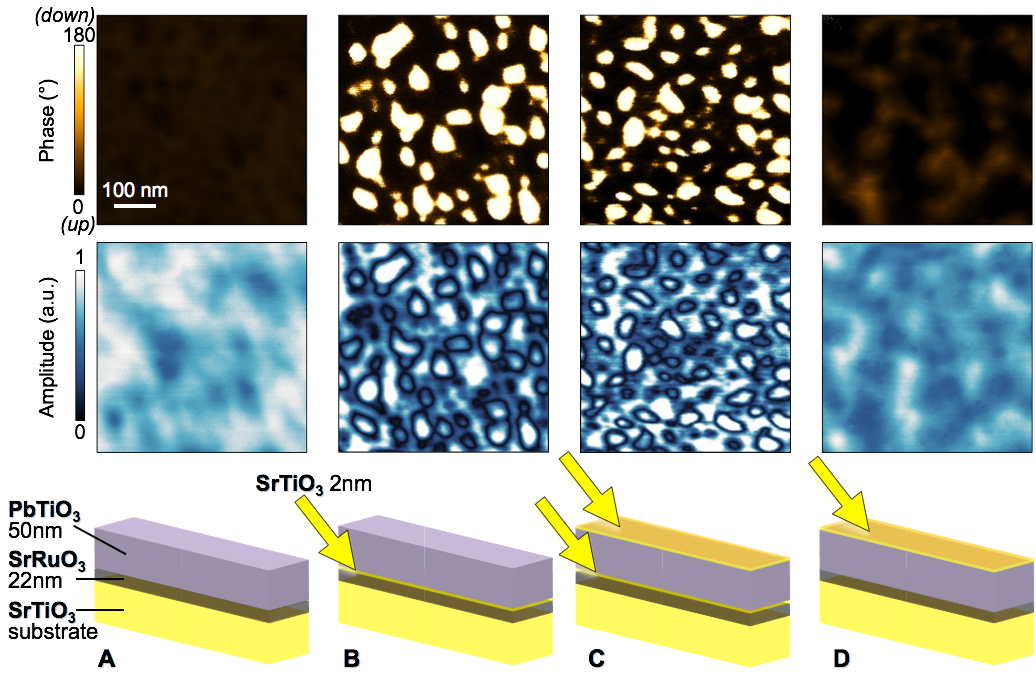}
\caption{\label{fig:PTO_differentSTO} Influence of the \sto\ spacers for 4 different samples with \pto\ thickness of 50 nm and (A) with no spacer, (B) with a bottom 2-nm \sto\ spacer only, (C) with both top and bottom 2-nm spacers, and (D) with only a top 2-nm spacer. All structures have a bottom \sro\ electrode of 22 nm. Top: phase images showing the local orientation of the polarization, uniformly {\it {\it up}} for A and D while with domains for B and C. Center: amplitude images for the 4 different samples, clearly showing domain walls (corresponding to a drop in the amplitude) for B and C. Bottom: schematic representation of the 4 corresponding samples.}
\end{figure}

To better understand the role played by the spacers, we studied several 50-nm-thick \pto\ films with different arrangements of 2-nm-thick \sto\ spacers as shown in \ref{fig:PTO_differentSTO}. The phase (top) and amplitude (center) signals were recorded over 500 x 500 nm$^2$ areas, applying the AC-voltage directly between the AFM tip and the bottom electrode.
The 50-nm \pto\ film grown directly on the \sro\ bottom electrode with no top or bottom spacer (sample A) is monodomain, with the polarization pointing {\it up}. 
This result is consistent with previous work by Nagarajan et al. \cite{Nagarajan-JAP-2006} who observed a monodomain {\it up} polarization state in \pzt\ films of similar thickness on SrRuO$_3$ electrodes. 
A monodomain state was also reported for 50 nm-thick \pto\ films with Nb-doped \sto\ and \lsmo\ electrodes \cite{Lichtensteiger-PRL-2005, Lichtensteiger-APL-2007} with {\it down} and {\it up} polarizations respectively. 
This preferential polarization direction implies the presence of a built-in field~\bibnote{The presence of a built-in field in our samples is also observed in switching spectroscopy PFM measurements. The phase and amplitude hysteresis loops acquired as a function of the DC voltage applied to the tip are systematically shifted, demonstrating the presence of a negative built-in bias, consistent with the observed preferential {\it up} polarization.}, which is most likely related to differences in band alignments at the interfaces of our asymmetric structure, though this was not investigated in detail in the present study. 

Adding a 2-nm-thick \sto\ spacer between the bottom electrode and a 50-nm-thick \pto\ film (sample B) completely changes its intrinsic polarization configuration from monodomain to polydomain as {\it down} polarized nanodomains form in the {\it up} polarized matrix. Note that the polarization direction of the matrix is the same as that of the monodomain film without the \sto\ spacers. In the amplitude image, the domains are outlined by a drastic drop of the piezoresponse at domain walls. The presence of the \sto\ spacer increases the distance between the screening charges from the bottom electrode and the ferroelectric polarization making the screening less effective and thus increasing the depolarization field inside the ferroelectric film~\cite{Mehta-JAP-1973,Wurfel-PRB-1973,Dawber-JPCM-2003,Lichtensteiger-Wiley-2011}. As a result, the polarization in the ferroelectric film is destabilized and domains of opposite polarization form in order to reduce the energy cost associated with this larger depolarization field. Therefore, without changing the thickness of the ferroelectric film, it is possible to tune the depolarization field and induce different domain configurations by adjusting the thickness of insulating spacers.

It is instructive to compare the change of the effective screening length induced by the \sto\ spacer with the intrinsic screening length of the metal-ferroelectric interface. The effective screening length $\lambda_{\rm eff,d}$ associated with a dielectric layer of thickness d and dielectric constant $\epsilon_{\rm d}$ is given by $\lambda_{\rm eff,d}={\rm d}/\epsilon_{\rm d}$. For a 2-nm-thick \sto\ layer with a room-temperature dielectric constant of 300\bibnote{The value measured in thin films and in \pto-\sto\ polydomain superlattices grown in the same conditions is close to 300.}, $\lambda_{\rm eff,d_{STO}}\approx$ 0.07 \AA, which is approximately half of the value of the intrinsic screening length for \pto-\sro\ and \sto-\sro\ interfaces (0.15 \AA) found using first-principles calculations~\cite{Stengel-Nature-2006,Stengel-NatMat-2009}. By adding an \sto\ spacer between the \sro\ electrode and the \pto\ film, the resulting effective screening length is the sum of the effective screening length of the \sto-\sro\ interface and that of the \sto\ spacer. The addition of the 2-nm \sto\ spacer is therefore expected to increase the screening length by roughly 50\% compared to the effective screening length of the bare \pto-\sro\ interface.

This simple argument, however, does not take into account the field-induced reduction of the \sto\ permittivity due to any built-in fields and fields induced by the ferroelectric layer, and therefore the simple calculation above may underestimate the increase in the $\lambda_{\rm eff,d_{STO}}$.

The effect of the top spacer, separating the ferroelectric film from the atmospheric adsorbates and the AFM tip, was also studied (samples C and D). The polarization pattern observed for the sample with both a top and a bottom spacer (\ref{fig:PTO_differentSTO}C) is very similar to that observed for the sample with only a bottom spacer (\ref{fig:PTO_differentSTO}B), with bubbles of {\it down} polarization embedded in a matrix of {\it up} polarization. On the contrary, the sample with only a top spacer (\ref{fig:PTO_differentSTO}D) is very similar to the sample without any spacer (\ref{fig:PTO_differentSTO}A), with a uniform {\it up} polarization. These observations indicate that the bottom spacer has more influence on the polarization configuration than the top spacer. Further investigation is needed to establish  the origin of this asymmetry.

\begin{figure}[!htb]
\includegraphics[width=0.9\columnwidth]{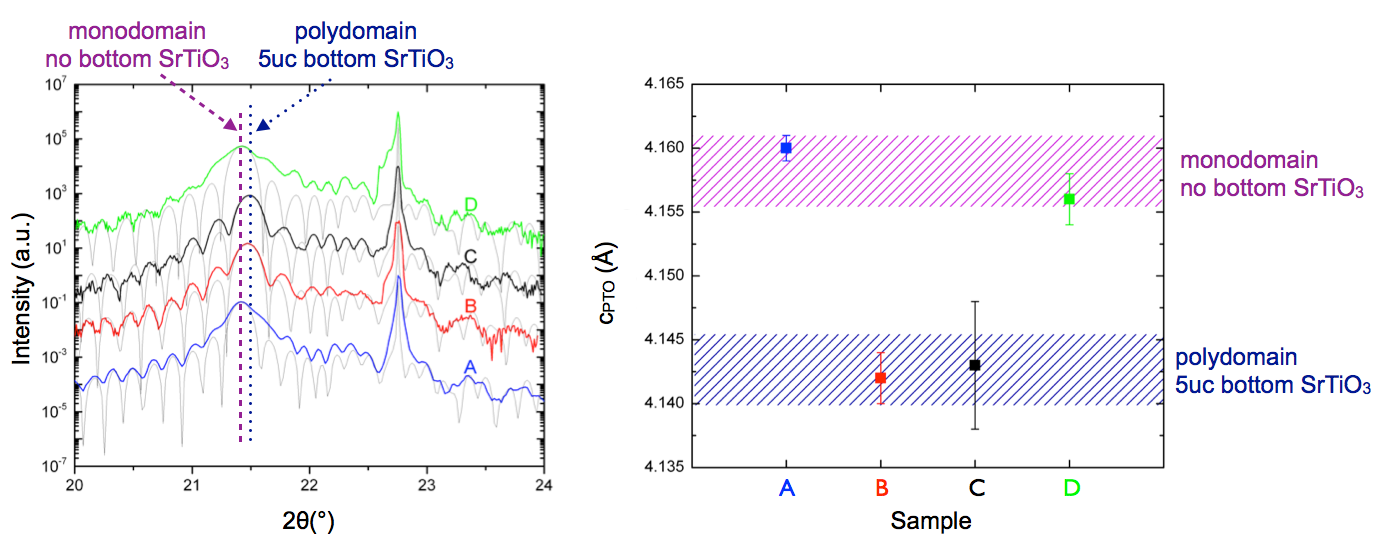}
\caption{\label{fig:cPTO_vs_STO} Influence of the \sto\ spacers on the lattice parameter of \pto\ for the 4 samples in \ref{fig:PTO_differentSTO}. Left: X-ray diffraction (XRD) intensities around the (001) reflection, together with the simulated intensities (shown in grey). Right: $c$-axis lattice parameter of \pto\ determined from the $(00l)$ specular reflections with $l=1,2,3$ and 4. Hatched regions are a guide to the eye. Samples A and D, without bottom \sto\ spacer, are monodomain, and display a larger lattice parameter than the polydomain samples B and C with a bottom \sto\ spacer.}
\end{figure}

\ref{fig:cPTO_vs_STO} shows the \pto\ $c$-axis values for the 4 samples shown in \ref{fig:PTO_differentSTO} obtained from X-ray diffraction measurements. Since all these samples have the same \pto\ thickness, the changes observed in their $c$-axis values are directly related to the different electrostatic environments. The  presence of the bottom \sto\  spacer has a clear effect on the $c$-axis lattice parameter with an average value of 4.158~\AA\ for the monodomain samples without a bottom spacer, and of 4.143~\AA\ for the polydomain samples with a bottom spacer. Our results are in line with the work of Takahashi~\etal~\cite{Takahashi-APL-2008}, who observe an increase of the $c$-axis lattice parameter for \pto\ films grown directly on Nb-doped \sto\ substrates when using photochemical switching to induce a transition from polydomain to monodomain. Again, the effect of the top spacer here is negligible.

\begin{figure}[!htb]
\includegraphics[width=\columnwidth]{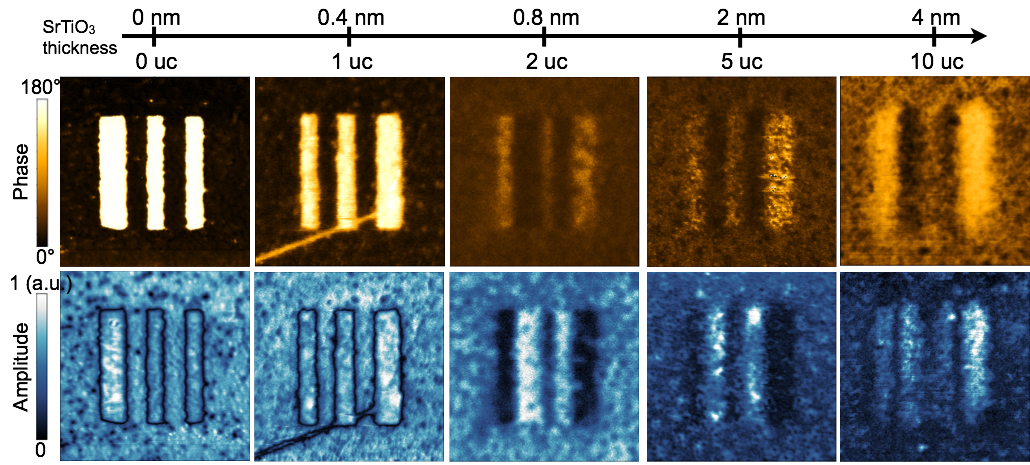}
\caption{\label{fig:20nmPTO_grid.png} PFM measurements for five different samples with 20-nm-thick \pto\ and top and bottom \sto\ spacers with thicknesses of 0, 1, 2, 5 and 10 uc. Phase (top) and amplitude (bottom) signals are shown on 1 x 1 $\mu$m$^2$ areas. The images were obtained after writing oppositely polarized regions by applying alternating $-$/+/$-$/+/$-$ DC voltages to the bottom electrode while scanning the grounded AFM tip over a 500 x 500 nm$^2$ region. These measurements demonstrate that the two samples with 0 or 1-uc-thick \sto\ spacer layers are monodomain, while samples with thicker \sto\ layers are polydomain.}
\end{figure}

To further investigate the effect of the \sto\ spacer thickness and test our capability to tune the depolarization field, we also grew a series of 20-nm \pto\ films with top and bottom \sto\ spacers of different thicknesses (0, 1, 2, 5 and 10 unit cells). For the samples with the thicker spacer layers, it was difficult to image directly the intrinsic domain configuration. Regions of opposite polarization were therefore written, by  applying alternating \mbox{$-$/+/$-$/+/$-$} DC voltages to the bottom electrode (typically $-$5V and +3V) while scanning the surface of the heterostructure with a grounded AFM tip. The resulting images are shown in~\ref{fig:20nmPTO_grid.png}.

For the samples with 0 and 1-uc-thick \sto\ spacers, only the lines written with negative voltage are visible in the phase image, with the domain walls clearly evident in the amplitude images. This observation points to monodomain samples with the as-grown polarization in the {\it up} direction. On the other hand, for the samples with 2, 5 and 10-uc-thick \sto\ spacers, the phase and amplitude images reveal a contrast for both types of written regions, which is the typical signature of polydomain samples with intrinsic domains too small to be clearly resolved using the AFM technique~\cite{Nagarajan-JAP-2006,Lichtensteiger-APL-2007}. The transition from monodomain to polydomain is consistent with the expected increase of the depolarization field as the \sto\ thickness increases.

\begin{figure}[!htb]
\includegraphics[width=0.75\columnwidth]{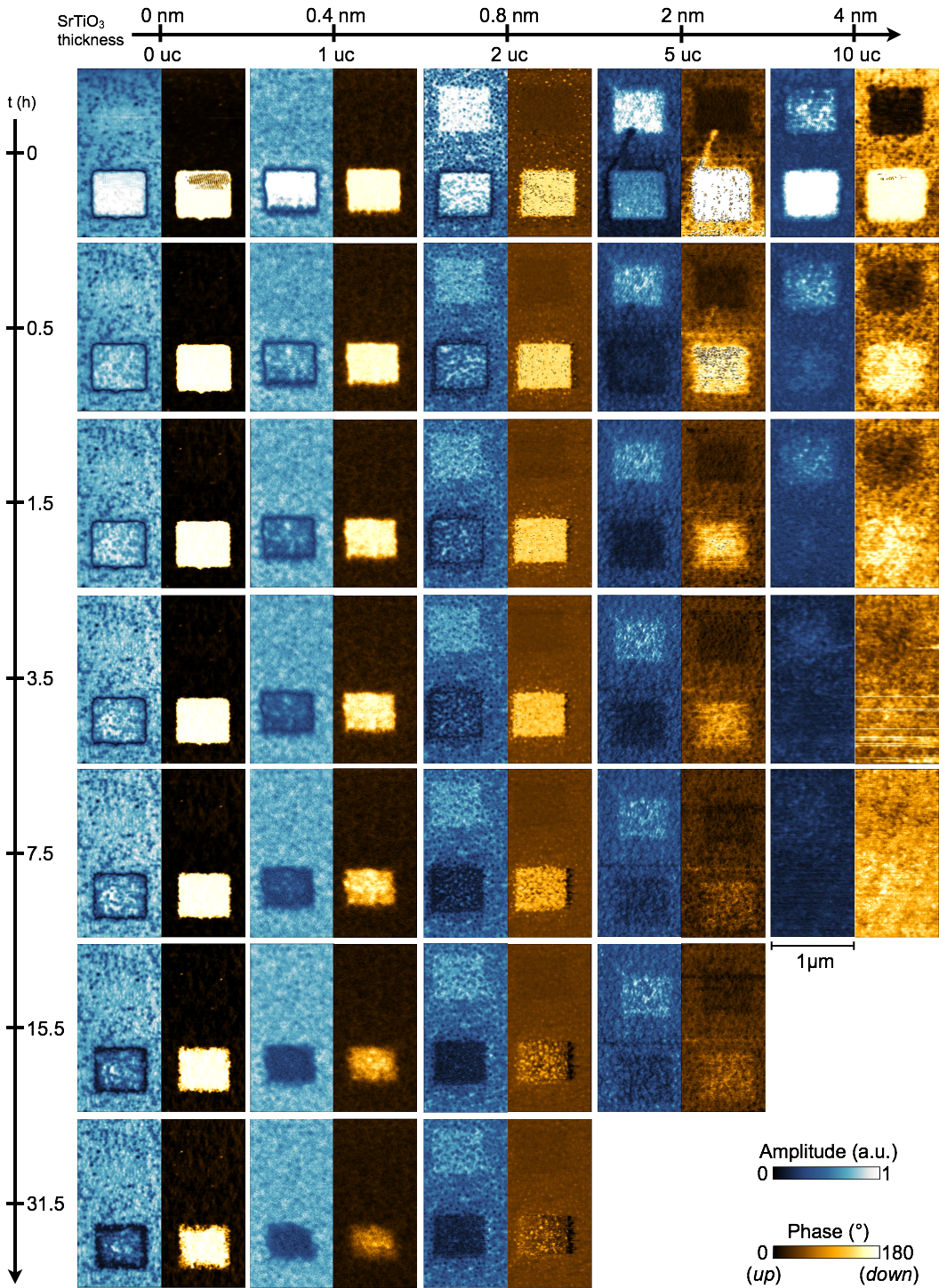}
\caption{\label{fig:Backswitching_20nmPTO} 
PFM measurements obtained on five different samples with 20-nm-thick \pto\ and top and bottom \sto\ spacers with thicknesses of 0, 1, 2, 5 and 10 uc. Amplitude (left) and phase (right) signals are shown for each sample on 2 x 1 $\mu$m$^2$ areas, at different times after writing two 500 x 500 nm$^2$ regions with {\it up} and {\it down} polarization. These measurements reveal different relaxation rates of the polarization for the different samples, with faster relaxation for samples with thicker \sto\ spacer layers.
}
\end{figure}

We also noticed that after writing regions of opposite polarization, the polarization relaxes at a different rate for different samples. We studied this relaxation by writing square regions 500 x 500 nm$^2$ with {\it up} and {\it down} polarization, and then imaged the written regions at different times after the writing. We were careful not to image the polarization state continuously, as the scanning itself was observed to influence the relaxation. \ref{fig:Backswitching_20nmPTO} shows the images obtained for the 5 samples mentioned above after different delay times.

For the two monodomain samples (0 and 1 uc), only the regions written with {\it down} polarization are visible in the monodomain {\it up} background. The written squares slowly disappear with time as the domain wall surrounding the written region gradually becomes rougher, reducing the size of the written region through domain wall motion driven predominantly by the built-in field.

For the sample with just 1 uc of \sto, in  addition to the roughening of the domain wall, we can notice that both the phase contrast and the amplitude decrease with time within the written region. This indicates that small {\it up} domains (below the resolution of the tip used) start to appear in the {\it down} written region, resulting in a decrease of the locally averaged phase and amplitude. In this latter case, the polarization relaxation is most probably due to the combined effect of the built-in field and a small depolarization field, which together exceed the threshold for nucleation of new domains.

For the polydomain samples, the polarization in the written squares relaxes back to its original polydomain state more rapidly for the sample with 5uc of \sto\ than for the one with 2uc of \sto , and even faster for the thicker \sto\ spacers (10uc). This polarization relaxation to a polydomain state is driven predominantly by the depolarization field~\cite{Kim-PRL-2005}. The different behaviors and time scales observed for the relaxation of the written domains indicate again that the depolarization field increases with the \sto\  layer thickness.

\begin{figure}[!htb]
\includegraphics[width=\columnwidth]{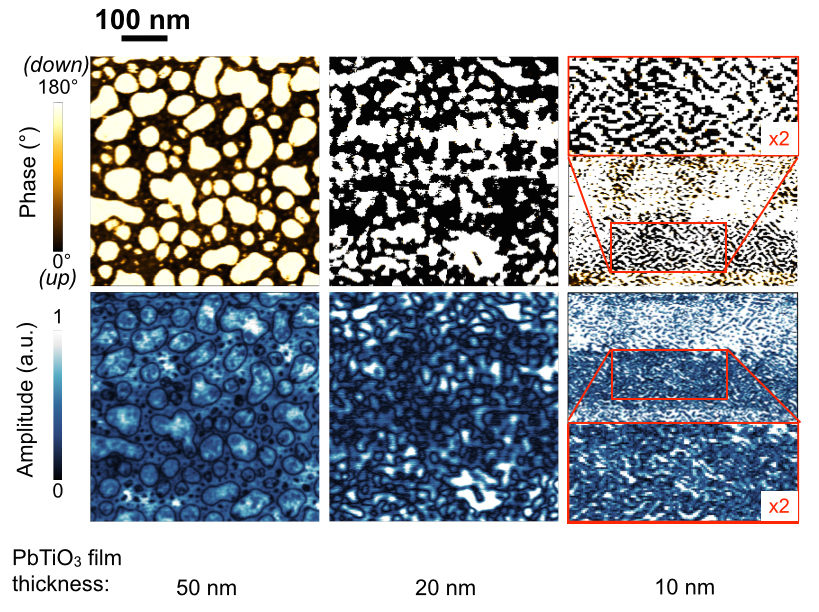}
\caption{\label{fig:PTO_2nmSTO} PFM measurements for three different samples with 50-nm-thick (left), 20-nm-thick (center) and 10-nm-thick (right) \pto\ layers. All samples have \sro\ bottom electrodes and top and bottom 2-nm-thick \sto\ spacers. Phase (top) and amplitude (bottom) signals are shown on 500 x 500 nm$^2$ areas. Part of the image obtained on the 10-nm-thick sample has been enlarged for clarity. From these measurements, we clearly see the decrease of the intrinsic domain size as the \pto\ film thickness decreases, as well as a change in shape from bubble-like for the thickest film to stripe-like for the thinnest one.}
\end{figure}

Introducing \sto\ spacer layers was key to obtain a polydomain configuration in samples that would otherwise be thick enough to stabilize a monodomain state. The next step was to study the polydomain configuration in samples with different ferroelectric film thicknesses.

\ref{fig:PTO_2nmSTO} shows PFM measurements obtained for three different samples with 50-nm (left), 20-nm (center) and 10-nm-thick \pto\ (right). All three samples have 2-nm-thick top and bottom \sto\ spacers. These measurements clearly demonstrate a decrease in the intrinsic domain size as well as a change in morphology as the film thickness is reduced: as the domain size decreases, the shape changes from bubble-like to stripe-like (or labyrinthine~\cite{Artemev-JAP-2008}). The domain pattern of the 10-nm-thick film resembles the stripe-like domains observed in 10-nm-thick \pto\ films grown on insulating \sto\ substrates using non-contact AFM~\cite{Thompson-APL-2008}.

The evolution of domain shape with film thickness is an interesting feature and we would like to speculate about its possible origins. A change of shape from stripes to bubbles upon domain period reduction is a rather general phenomenon that has been observed in a number of different systems including magnetic garnet, Langmuir films and ferrofluids~\cite{Seul-SCI-1991,Elias-JdPI-1997}. For magnetic films in particular, the stripe-to-bubble transition has been extensively studied, both theoretically and experimentally, in the context of bubble memories~\cite{Bobeck-IEEE-1975}. Theoretical work has shown the energies of both configurations to be very similar, but in the absence of applied field the ground state is always the striped phase, which most effectively compensates the demagnetising field~\cite{Cape-JAP-1971}. As the applied field increases above a critical value, however, the energy balance shifts in favor of a bubble domain state. Similar field-induced splitting of stripes into bubbles in ferroelectrics has also been studied theoretically by Chensky and Tarasenko~\cite{Chenskii-ZET-1982} and more recently in the groups of Bellaiche~\cite{Lai-PRL-2006,Kornev-PRL-2004} and Roytburd~\cite{Artemev-JAP-2008}, while Luk'yanchuk \etal ~\cite{Lukyanchuk-arXiv-2013} investigated the stability of individual ferroelectric bubble domains. One may therefore expect that the domain shape will, at least in part, be controlled by the competition between the external field, which favours bubble domains, and the depolarization field, which favours stripe-domain formation. In our case, the external field comes from the built-in bias.

As the depolarization field increases with decreasing film thickness, it favours the striped phase over the bubble phase \cite{Kornev-PRL-2004}. In thicker films, the effect of the depolarization field is weaker and the built-in bias may favour bubble domains. The film-thickness dependence of the built-in field in our samples, however, is currently less well understood as the precise origin of this field remains to be investigated. Previous studies \cite{Tagantsev-IntFerro-1995,Ng-PRB-2012} have shown that in the presence of space-charge layers this field may increase or decrease with film thickness depending on whether the film is fully or partially depleted and therefore a quantitative assessment of this scenario is not currently possible.

In addition, the above electrostatic considerations are not sufficient to explain all the features of the data. In particular, the highly irregular shapes of the domains, as well as the internal structure in the amplitude images, indicate that defect-induced fluctuations in the polarization profile also play an important role. This is supported by our observations that in the 50-nm-thick films the location of the bubble domains is predetermined and the domains always reappear in the same places after the films are electrically poled into a monodomain state. The defect structure may also play an important role in the stripe-to-bubble transition itself by breaking the long-range coherence of the striped structure. Thus, while the observed evolution of domain size and shape might be qualitatively captured by the simple electrostatic considerations above, the precise domain morphology is likely to be governed by a more complex combination of the effects of depolarization, defect-induced pinning and nucleation, and additional screening due to free carriers within the films.

Phase-field calculations\cite{Chen-AnnuRev-2002} that allow the inclusion of realistic materials parameters and extend previous phenomonelogical models~\cite{Kopal-Ferroelectrics-1999,Bratkovsky-PRL-2000,Lukyanchuk-PRL-2009} to more complex domain morphologies would be particularly helpful in disentangling the role of depolarization effects from that of extrinsic factors such as disorder~\cite{Giamarchi-2006}.

\begin{figure}[!htb]
\includegraphics[width=0.9\columnwidth]{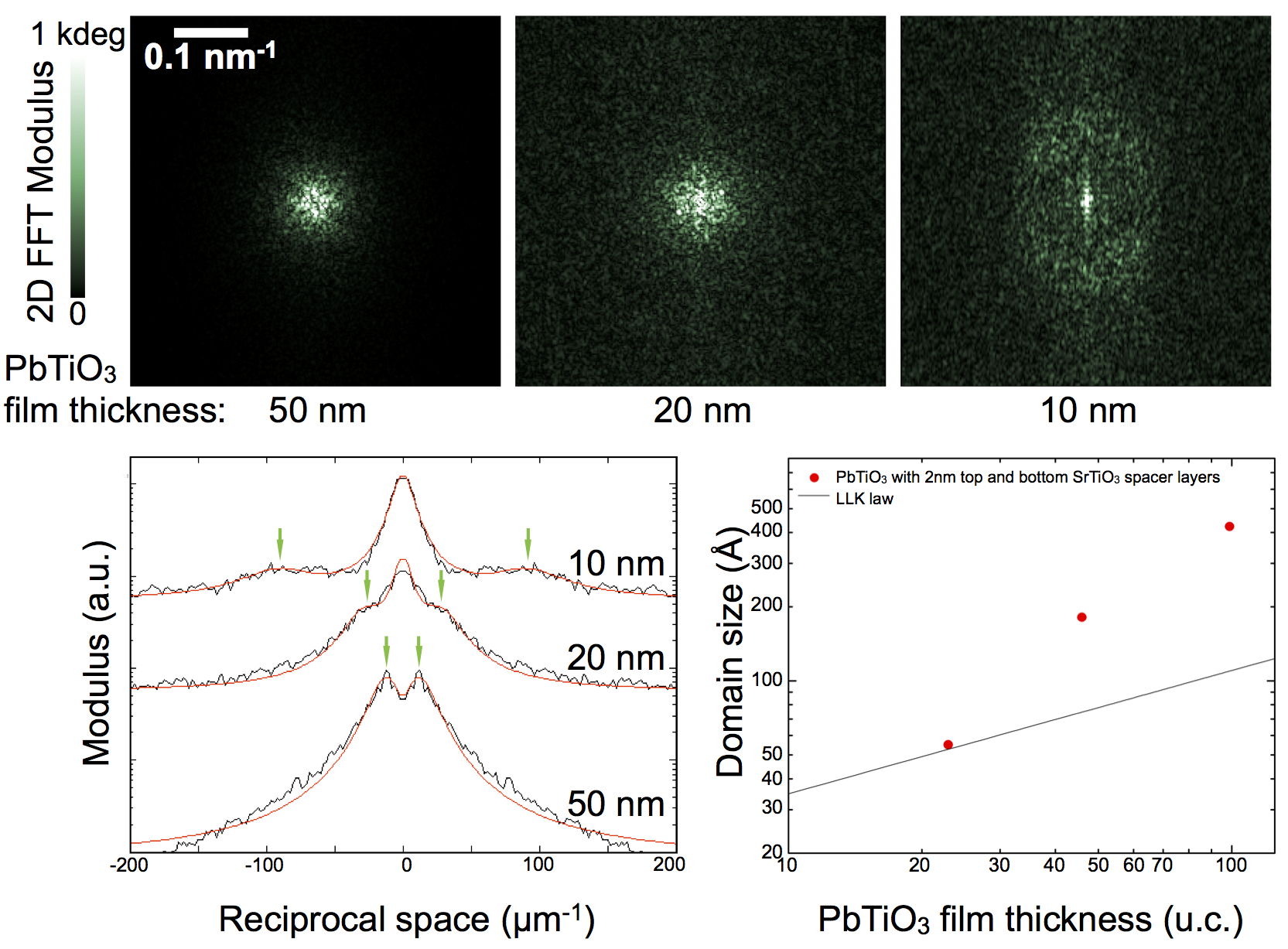}
\caption{\label{fig:DomainsSizes} 2D FFT transforms of the raw PFM phase images (shown in~\ref{fig:PTO_2nmSTO}) were used to estimate the domains sizes. (Top) 2D FFT Modulus of the PFM phase images of the domains for the 50-nm, 20-nm and 10-nm-thick films. (Bottom left) Symmetrised line profiles obtained from the radial average of the 2D FFT Modulus, showing the presence of satellite peaks from which the domain sizes were extracted. Lorentzian functions were used to fit each profile, and their center was used to determine the domain sizes. (Bottom right) The domain sizes are plotted as a function of the \pto\ film thickness (red dots). For comparison, the black line shows the Landau-Lifshitz-Kittel scaling of domains in PbTiO$_3$ films studied by Streiffer \etal \cite{Streiffer-PRL-2002}.}
\end{figure}

To estimate the domain sizes for the different film thicknesses, we have Fourier transformed the raw PFM phase images from \ref{fig:PTO_2nmSTO} using Gwyddion 2D-FFT. The obtained reciprocal space distributions of the relevant length scales are shown in~\ref{fig:DomainsSizes}.

The reciprocal space image for the 10-nm film shows a ring-like satellite structure around the central peak corresponding to a real-space domain size of approximately 5 nm. This value is in excellent agreement with the values obtained using X-ray diffraction measurements on \pto -\sto\ superlattices with electrostatically decoupled \pto\ layers of similar thickness~\cite{Zubko-NanoLetters-2012}. It fits in well with the Landau-Lifshitz-Kittel (LLK)~\cite{Landau-PZS-1935, Kittel-PR-1946, Kittel-RMP-1949} scaling  of the stripe-domains observed in PbTiO$_3$ thin films  by Streiffer \etal ~\cite{Streiffer-PRL-2002} and shown in \ref{fig:DomainsSizes}. This implies that for the 10-nm film there is little interaction of the domain structure with the metallic electrode and atmospheric screening charges. For thicker films, the FFT pattern is less isotropic but nevertheless allows rough estimates of the characteristic length scales to be determined as 18 nm for the 20-nm-thick film and 42 nm for the 50-nm-thick film. These sizes significantly exceed the LLK expectation. Deviations from the square-root LLK law have been reported by several groups and can arise for a number of reasons (see Ref.~\cite{Catalan-PRL-2008,Tagantsev-Springer-2010,Nesterov-APL-2013,Scott-JPCM-2014} and references therein). In our case, it is most likely due to the enhanced electrostatic coupling between the ferroelectric and metallic layers mediated by the stronger electric fields induced in the \sto\ layers by the larger domain structures~\cite{Kopal-Ferroelectrics-1997}. 

In conclusion, by introducing \sto\ spacer layers, we were able to tune the effective screening length and control the depolarization field in ferroelectric heterostructures. This, in turn, was exploited to induce a polydomain state in otherwise monodomain samples. The simple method used here to control the depolarization-field-induced domain structure offers not only an excellent tool for fundamental studies of ferroelectric nanodomains and polarization stability in ferroelectrics, but may also be used to enhance the dielectric properties of ferroelectric thin films by exploiting the large domain wall contributions to the dielectric permittivity~\cite{Zubko-PRL-2010} or to engineer materials whose properties are dominated by the exotic functionalities that have recently been discovered at ferroic domain walls~\cite{Catalan-RevModPhys-2012}.

\begin{acknowledgement}
The authors thank P. Aguado Puente, C. Blaser, I. Gaponenko, J. Guyonnet, J. Junquera, P. Paruch and B. Ziegler for helpful discussions, M. Lopes and S. Muller for technical support, and acknowledge funding from the Swiss National Science Foundation through the NCCR MaNEP and division II, and the EU project OxIDes.
\end{acknowledgement}

\providecommand*\mcitethebibliography{\thebibliography}
\csname @ifundefined\endcsname{endmcitethebibliography}
  {\let\endmcitethebibliography\endthebibliography}{}


\begin{mcitethebibliography}{55}
\bibitem[Not()]{Note-1}
SrRuO$_3$ was deposited at 640$^\circ$C in 100 mTorr of O$_2$/Ar mixture of
  ratio 3:60 using a power of 80 W. The \pto\ and \sto\ layers were deposited
  at 530$^\circ$C in 180 mTorr of 29:20 O$_2$/Ar mixture using a power of 60 W.
  A Pb$_{1.1}$TiO$_3$ target with a 10\% excess of Pb was used to compensate
  for the Pb volatility, while stoichiometric targets were used for \sro\ and
  \sto .\relax
\mciteBstWouldAddEndPunctfalse
\mciteSetBstMidEndSepPunct{\mcitedefaultmidpunct}
{}{\mcitedefaultseppunct}\relax
\EndOfBibitem
\bibitem[Not()]{Note-3}
The value measured in thin films and in \pto-\sto\ polydomain superlattices
  grown in the same conditions is close to 300.\relax
\mciteBstWouldAddEndPunctfalse
\mciteSetBstMidEndSepPunct{\mcitedefaultmidpunct}
{}{\mcitedefaultseppunct}\relax
\EndOfBibitem
\end{mcitethebibliography}
\end{document}